# Optical spin Hall effect pattern switching in polariton condensates in organic single-crystal microbelts

Jiahuan Ren,[1,2] Teng Long,[1] Chunling Gu,[3] Hongbing Fu,[1] Dmitry Solnyshkov,[4,5,*] Guillaume Malpuech,[4] Qing Liao[1,*]

[1] Beijing Key Laboratory for Optical Materials and Photonic Devices, Department of Chemistry, Capital Normal University, Beijing 100048, China

[2] College of Physics Science and Technology, Hebei University, Baoding 071002, China

[3] Institute of Process Engineering, Chinese Academy of Sciences, Beijing, 100190, China

[4] Institut Pascal, PHOTON-N2, Université Clermont Auvergne, CNRS, Clermont INP, F-63000 Clermont-Ferrand, France

[5] Institut Universitaire de France (IUF), 75231 Paris, France

* marks the corresponding authors

Abstract

Topological polaritons, combining the robustness of the topological protected edge states to defects and disorder with the strong nonlinear properties of polariton bosons, represent an excellent platform to investigate novel photonic topological phases. In this work, we demonstrated the optical spin Hall effect (OSHE) and its symmetry switching in the exciton-polariton regime of pure DPAVBi crystals. Benefiting from the photonic Rashba-Dresselhaus spin-orbit coupling in organic crystals, we observed the separation of left- and right-circularly-polarized polariton emission in two-dimensional momentum space and real space, a signature of the OSHE. Above the lasing threshold, the OSHE pattern changes due to transverse quantization in the microbelt. This device without superlattice structure has great potential applications in topological polaritons, such as information transmission, photonic integrated chips and quantum information.

**Introduction**

The investigation of topological phenomena in condensed matter[1,2] has become a mainline of scientific activity in the past decades because the topological robustness makes topological states immune to perturbations and allows unidirectional transmission without backscattering. Extending this idea to the realm of photonics was initially suggested by Haldane and Raghu[3], and then an analog of the quantum anomalous Hall effect has been realized in the microwave range through breaking the time-reversal symmetry (TRS)[4]. Subsequently, many effects have been demonstrated in topological photonics, including quantum optics[5], topological lasing[6] and non-Hermitian topological photonics[7], which arouses considerable upsurge of engineering photonic band structures and related topological photonic devices[8].

One of the key approaches in topological photonics is based on the use of the inherent topology of the photonic modes linked with their transverse nature. Other approaches can be based on the sublattice pseudospin and are not specific to photonics. The natural photonic spin-orbit coupling (SOC) is called "the transverse electric and transverse magnetic (TE-TM) splitting". The foundations for topological photonics were laid by the discovery of the optical spin Hall effect[9,10] (OSHE), also called the (spin) Hall effect of light[11]. It is a direct consequence of the TE-TM SOC, subject of extended studies, and its different versions have already found practical applications[12]. The patterns generated by OSHE depend on the symmetry of the SOC[13] and allow distinguishing, for example, the TE-TM (double winding) from the Rashba SOC[14] (single winding).

In addition to the TE-TM SOC, structures with distributed Berry curvature require[15]

TRS breaking by applied magnetic field via the Faraday effect or chiral symmetry breaking by optical activity. However, the magneto-optical response that allows to break TRS in classical materials is very weak at optical frequencies. An alternative solution is to use the exciton-polariton (polariton) concept, in which the strong coupling between photons and excitons forms half-light, half-matter quasiparticles[16]. In this strong-coupling system, a nontrivial photon topology can be created with the help of electronic degrees of freedom[17,18]. On the other hand, as recently discovered[19], the chiral symmetry in cavities can be broken by the resonance of the modes of opposite parity, which can be possible thanks to large birefringence. Controllable birefringence can be induced in liquid crystal microcavities, where it has already allowed to observe tunable OSHE[20]. It can also be achieved by selective strong coupling in organic systems. This effect gives rise to Rashba-Dresselhaus (RD) SOC (emergent optical activity)[21]. Organic molecular crystals were demonstrated as ideal materials for room-temperature polariton and polariton condensation due to high binding energy and strong oscillation strength of their tightly bound Frenkel excitons[22]. Recently, the pseudospin valley Hall effect had been found in organic crystal-filled cavities based on the crystalline anisotropy-induced photonic SOC without magnetic field or honeycomb lattices[21]. This gives an opportunity for exploring new topological polariton matter that bridges photonic and electronic systems at room temperature.

Compared with bare photonic systems, the excitonic component in polaritons introduces strong interactions between particles and strong nonlinearities that can be useful for studying nonlinear topology[23] and actively controlling the spin and valley degrees of freedom in the future devices[6]. Importantly, polaritons can exhibit Bose-Einstein condensation (BEC) because they are composite bosons which allows them to macroscopically occupy the ground state[24] or at another state. This also permits to achieve polariton lasing without the need for population inversion. To date, polariton BEC and related phenomena such as polariton topological insulators[25], topological polariton lasers[23,26], topological $Z_2$ exciton-polaritons[27] have been demonstrated both theoretically and experimentally[18,28]. In these works, high-quality microcavities based on two distributed Bragg reflectors (DBRs)[22] were used for constructing topological polaritonic states. Such device configuration inevitably brings about the complexity of manufacturing process, especially in the visible-light range, which might reduce the integration and practicality of the devices.

In this work, we have observed polarization-selective strong coupling, polariton formation and condensation in pure organic microscale crystals without any mirrors. The smooth and flat lateral sides of the organic single crystal allow to build a natural resonant cavity, which provides strong exciton-photon coupling. Here, the polariton losses resulting from the low quality factor of the natural resonant cavity can be compensated by the increase of the densities of Frenkel excitons through selecting a suitable organic semiconductor with high optical transition strength[29] and low non-radiative losses[30]. We measure the optical spin Hall effect patterns in real and reciprocal space both below and above threshold, demonstrating their symmetry switching. Below threshold, the 2-fold pattern is a signature of the RD SOC (optical activity). Above threshold, the coherence length increases and the in-plane quantization of the modes of the microbelt becomes important. We observe an interference pattern stemming from two quantized modes, with a 4-fold symmetry due to the TE-TM SOC. Our findings reveal the superiority of organic single-crystals in view of the fabrication of mirrorless polariton-based spintronic devices for on-chip integrated photonic circuit applications.

**Results and discussion**

The 4,4′-bis[4-(di-p-tolylamino)styryl]biphenyl (DPAVBi) microcrystals were prepared via the solution reprecipitation method[31]. The details on the synthesis of DPAVBi molecules, the preparation of microbelts and their characterization are provided in Supplementary Materials (see Materials and Methods and Figure S1)

Since the interaction strength of light and matter is described by the scalar product of the transition dipole-moment (μ) and the cavity electric field[32], the molecular orientation plays an important role for the formation of strong coupling between photons and excitons. Upon optical excitation, photons strongly couple with Frenkel excitons to generates polaritons in the DPAVBi microbelts (Figure 1a). The Fabry-Pérot microcavity formed by the two lateral sides in the width direction of the microbelt provides optical confinement to enhance the light-matter interaction (black double arrow line in Figure 1a). Polaritons with different wave vectors leak out of the microbelt (as shown in the schematic diagram of Figure 1a) and are collected by the angle-resolved photoluminescence (APPL) measurement. In the process of ARPL measurements, the microbelt length direction was always parallel to the slit of the spectrometer (Figure 1b).

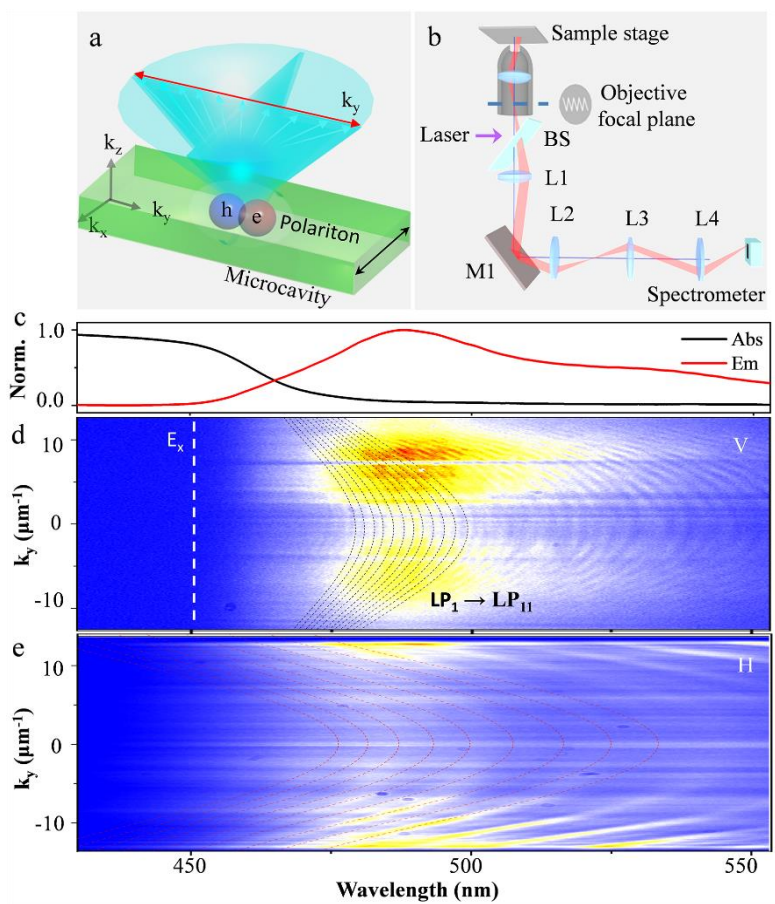

**Figure 1. a**: Schematic of pure DPAVBi crystal. **b**: Schematic experiment setup. **c**: The absorption and emission spectra of DPAVBi crystals. **d**, **e**: The V (parallel to the crystal long axis) and H (vertical to the crystal long axis) polarized angle-resolved PL spectra of the pure DPAVBi crystal excited by the 405-nm continuous wave laser. Note: Figure 1e shows a 10x magnification of the data. $E_x$ is the position of the exciton energy. The black and red dashed lines are attributed to the lower polariton branches ($LP_1$ to $LP_{11}$) and uncoupled cavity photon modes, respectively.

Figure 1c shows the absorption and emission spectra of the single DPAVBi microbelt on a quartz. The absorption spectrum (black line) exhibits a broad absorption band with a maximum at 452 nm is observed. Due to the H-aggregates characteristics of the forbidden 0-0 emission transition and allowed 0-n (n>1) sideband emissions[33], DPAVBi crystals show a strong PL emission with 0-1 and 0-2 emissive bands at 494 nm and 528 nm, respectively (red line). We performed polarization-dependent ARPL of the selected DPAVBi microbelt by adding a linear polarizer in the detection optical path. Figures 1d and 1e present the vertical (V)-polarized (parallel to the crystal long axis) and horizontal (H)-polarized (vertical to the crystal long axis) ARPL spectra which reveal the relation between the wavelength and the in-plane wavevector $k_y$, respectively. Two sets of orthogonally linearly polarized modes with distinctive curvature are observed. The H-polarized modes with larger curvature are distributed evenly in the spectral range (460-550 nm), which can be perfectly fitted by the planar (Fabry-Pérot) cavity modes (red lines in Figure 1e). The fitted effective refractive index ($n_{eff}$) is determined to be 1.60. In sharp contrast, the dispersions of V polarization show small curvatures (Figure 1d), which suggests that there might be strong coupling between light and matter here. To determine $n_{eff}$ for V-polarized modes, we first simulate those modes in the spectral range (520-550 nm) because their photon component is close to cavity-photon modes. The fitted $n_{eff}$ is determined to be 2.15 (also see Figure S2). Notably, the modes with smaller curvature distribution appear at the spectral range of 460-500 nm near the exciton resonance at 452 nm (white line in Figure 1d). These small-curvature modes are in good agreement with the lower polariton ($LP_1$-$LP_{11}$) dispersions calculated by using the coupled harmonic oscillator model[30] (black lines in Figure 1d). This clearly indicates that the strong coupling between exciton and photons occurs in pure DPAVBi microbelt. The fitting parameters for $LP_n$ (n = 1 to 11) at θ = 0° are summarized in Table S1 and the Rabi splitting energy (Ω) is calculated to be 390 meV. It can be concluded that the strong coupling only occurs in the V polarization, because of the much stronger excitonic absorption for V-polarized light than for H-polarized one[6].

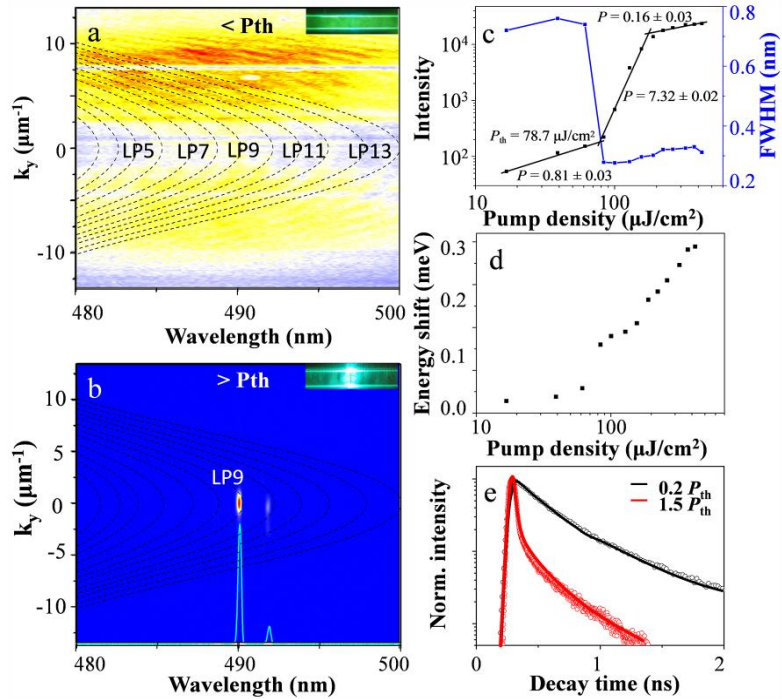

**Figure 2. a,b**: Angle-resolved PL measured at P < $P_{th}$ (**a**) and P > $P_{th}$ (**b**). Insets of (**a**) and (**b**) are the corresponding real-space images of the DPAVBi crystal. **c**: The intensity (black squares) and the line width (blue squares) of the LPB as a function of pump density. **d**: The blueshift energy of the $LP_9$ as a function of the pump density. **e**: The PL decay time under different pump densities.

We subsequently pumped the DPAVBi microbelt through off-resonance excitation by focusing (with a 100 × 0.95 NA objective) the second harmonic (λ = 400 nm, pulse width 150 femtosecond (fs)) of a Ti:sapphire regenerative amplifier to a 40 μm diameter spot with a Gaussian beam profile. At the low pumped density, the DPAVBi microbelt shows uniform green-color emission according to the PL micrograph (inset of Figure 2a). ARPL emission from the large wavevector positions are clearly observed, which follow the simulated polariton dispersions as indicated by black dotted lines (Figure 2a), proving that the PL signal indeed originates from polariton dispersion branches. With the increase of pumping density to $P$ = 138 μJ/cm², the polaritons condense at the bottom of the $LP_9$ near the $k_y$ = 0 (Figure 2b). This way the polaritons relax to the bottom of $LP_9$ branch by stimulated scattering and form polariton condensates. The cyan solid line in Figure 3b presents the corresponding PL spectrum of the LP at θ = 0°, which exhibits a maximum at 490 nm with the full width at half-maximum (FWHM) of about 0.7 nm. These results confirm the macroscopic ground state population of the LP branches, one of the main characteristics of polariton condensation[34]. Figure 2c presents the integrated intensity (black squares) and FWHM (blue squares) of the LP emission as a function of the pump density, giving rise to an "S" shaped curve. The intensity dependence for the $LP_9$ branch is separately fitted to power laws $x^p$ with $p$ = 0.81 ± 0.03, 7.32 ± 0.02, and 0.16 ± 0.03, respectively. Therefore, the condensation threshold is identified as $P_{th}$ = 78.7 µJ/cm² between the sublinear and superlinear regions. The FWHM of the $LP_9$ is drastically narrowed from 0.76 nm below the threshold to 0.27 nm above the threshold. Meanwhile, a clear power-dependent blueshift of polariton energy (Figure 2d) beyond the $P_{th}$, resulting from the interaction between polaritons[35], evidences that the system remains in the strong coupling regime. Figure 2e presents the time-resolved PL from the $LP_9$ branch measured with a streak camera. At a low pump density of 0.2 $P_{th}$, the polariton PL follows a single-exponential decay with a lifetime of $\tau$ = 0.29 ± 0.01 ns. Upon increasing the pump density to above the threshold of 1.5 $P_{th}$, the PL decay time collapses to < 15 ps and is limited by the resolution of our apparatus, indicating the stimulated scattering to the condensate. The nonlinear increase of the integrated PL intensity, the drastically narrowing of the FWHM, and the collapse of the emission lifetime indicate the polariton condensation.

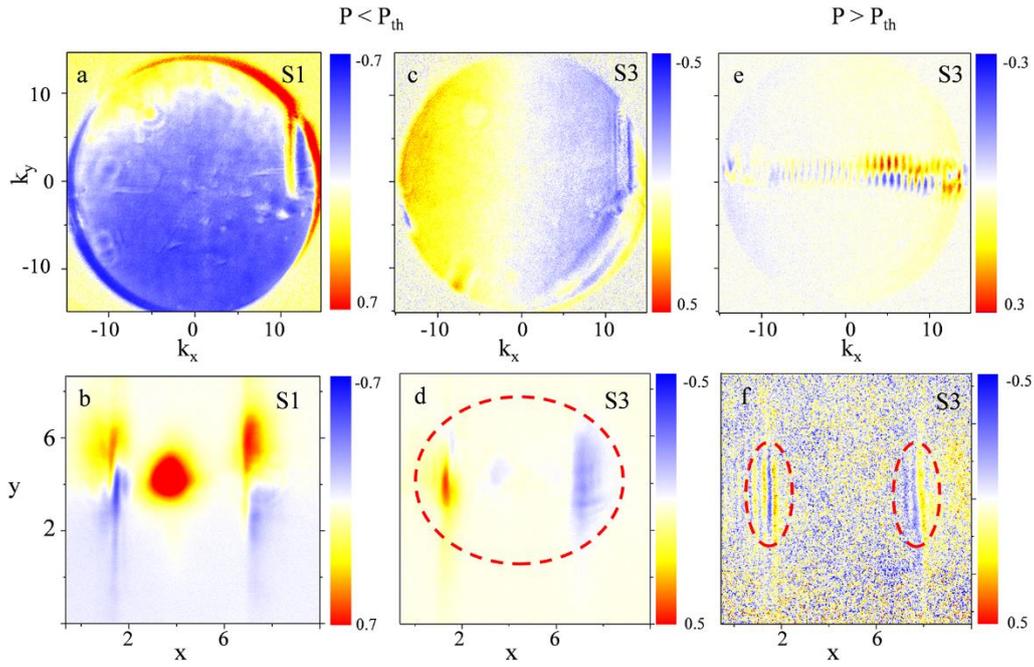

**Figure 3.** The Stokes parameters S1(**a,b**) and S3(**c,e,d,f**) of PL image in the 2D momentum space (top) and real space (bottom) from the pure DPAVBi microbelt at $P < P_{th}$ (a,b,c,d) and $P > P_{th}$ (e,f).

Different SOCs in the microcavities can generate effective magnetic fields acting on the cavity photons[36], which may lead to the separation of photon spins and produce the optical spin Hall effect[9]. To map this spin separation, we measured 2D wavevector of the Stokes vector components for momentum space and real space from polarization-resolved PL of the DPAVBi microbelt (Figure 3). Here, the 405-nm continuous-wave laser is used to excite the DPAVBi microbelt. Surprisingly, the Stokes component $S1$ shows that the linear polarizations are separated along the $k_y$-direction in 2D momentum space (Figure 3a). In the image of 2D real space (Figure 3b), the component $S1$ of the Stokes vector cancel and change sign around the positions of near y = 4 in the y-direction.

It is important to note that the OSHE as spatial spin separation may arise in two regimes. In the first case[9], one excites a superposition of several eigenstates, and their beatings lead to spin precession and spatial separation during the propagation. In the second case one excites the eigenstates of the system (e.g. only the lowest branch in a 2-band model). These eigenstates for opposite wave vectors may have opposite spins and opposite group velocities, which again leads to their spatial separation during the propagation.

Our observations below threshold correspond to the second case: no coherent superposition and precession is possible under spontaneous scattering. This scattering populates the eigenstates of the system differently, depending on their energy, which leads to spin imbalance for each wave vector. The measured Stokes vector $S3$ exhibits an obvious spin separation, i.e., the left and right circularly-polarized separation along the $k_x$ axis (in Figure 3c). In the 2D real space (Figure 3d), the Stokes vector $S3$ component is maximal at the edges of the sample. These observations mean that the left and right circularly polarized components and propagate in opposite directions along the width direction of the microbelt. This phenomenon of real- and reciprocal-space separation can be seen as unidirectional propagation of opposite pseudospins.

Furthermore, we also measured the Stokes vector $S3$ component of the 2D wavevector in momentum space and real space above $P_{th}$ by the 400-nm fs laser. From the 2D momentum space (Figure 3e), the polariton condensate exhibits a four-fold pattern typical for TE-TM SOC, which results from the spin precession due to the linear polarization of the polariton condensate (only H-branches are strongly coupled), which corresponds to a superposition of TE and TM eigenstates. The RD SOC responsible for opposite circular polarizations along $k_x$ cannot play a role any more due to the increased coherence length of the system: the reflection on the edges flips the wave vector without flipping the spin, and thus the S3 component averages out for $k_y$=0. The interference fringes observed in the k-space confirm the spatial quantization of the modes across the microbelt. The role of the RDSOC is therefore suppressed, and the TE-TM becomes dominant, which explains the observed four-fold OSHE pattern.

We also measured the dispersion relationship between wavelength and $k_y$ wavevector ($k_x$ = 0) of the circularly polarized light at the pump density above $P_{th}$ (Figure S3). The results clearly indicate that the left and right circular polarizations are splitting along the $k_y$ direction at the wavelength of 490 nm. The image of 2D real space (Figure 3f) also shows the spin separation along the y-axis, i.e, the length direction of the DPAVBi microbelt. All this, together with the interference fringes from spatial quantization mentioned above, is consistent with simulation results (Figure 4). Thus, the unidirectional propagation of left-right circularly polarized light over spatially quantized states is achieved in the DPAVBi crystals.

To qualitatively reproduce the results of the experiment, we use the spinor generalized Gross-Pitaevskii equation implementing a two-band model:

$$i\hbar \frac{\partial \psi_\pm}{\partial t} = (1-i\Lambda)\hat{T}_\pm \psi_\pm + \beta \left( i\frac{\partial}{\partial x} - \frac{\partial}{\partial y} \right)^2 \psi_\mp + i\gamma(n_{tot})\psi_\pm + U\psi_\pm,$$

where $\hat{T}_\pm$ is the kinetic energy operator accounting for several effects: the dispersion relation of polaritons, the gauge field contribution of the RD SOC, and the energy relaxation mechanisms; $\beta$ is the spin-orbit coupling constant describing the TE-TM splitting; $\gamma(n)$ is the net gain determined by the losses of the mirrorless cavity, non-radiative losses, and saturated gain from the excitonic reservoir, whose spatial profile is determined by the experimental conditions; $U$ is the confining potential determined by the in-plane width of the microbelt. The two bands implemented in the model represent the pair of bands that exhibit condensation in experiment and necessarily have the largest population below threshold. The equation is solved using the 3$^{rd}$-order Adams-Bashforth method with the kinetic energy and TE-TM operators calculated using the Fast Fourier Transform accelerated by the Graphics Processing Unit. The results of the simulations are shown in Figure 4.

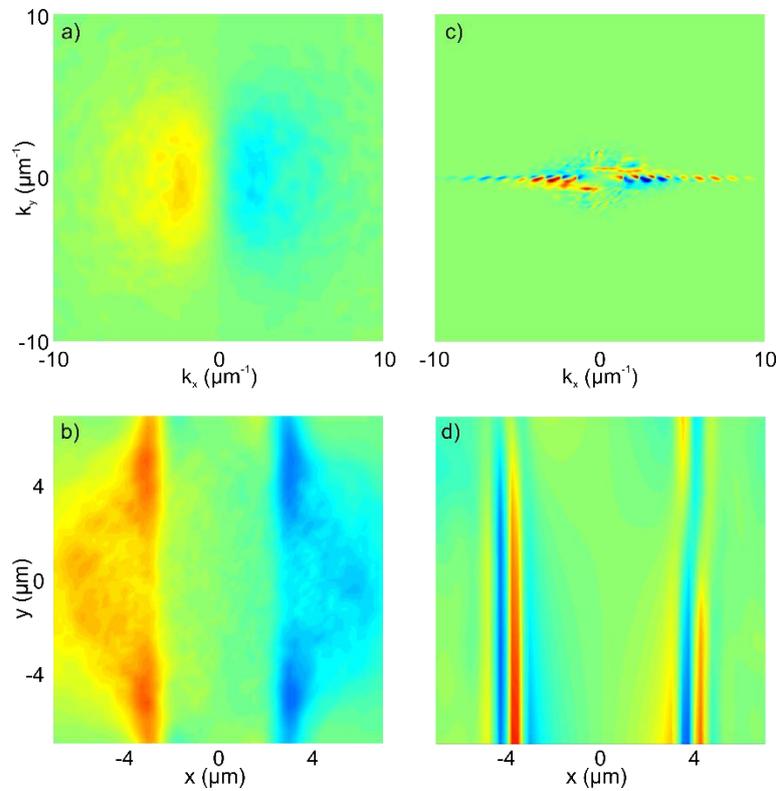

**Figure 4.** Numerical simulations showing the calculated $S3$ Stokes parameter: (**a,b**) below threshold, (**c,d**) above threshold. Top row (**a,c**) — momentum space, bottom row (**b,d**) — real space.

Below threshold (Figure 4a,b), the spatial quantization of the modes across the microbelt is obscured by thermal broadening. The results are therefore similar to the localized excitation of a planar cavity with a particular spin-orbit coupling: the OSHE. The two-lobe symmetry of the observed $S3$ pattern (Figure 4a) confirms the leading role of the RD SOC, which dominates for the wave vectors involved. Real-space emission (Figure 4b) is in agreement with this picture: particles coming to the edge of the sample have to propagate in a given direction and thus exhibit the same circular polarization as in Figure 4a.

In the condensed regime (Fig. 4c,d) the linewidth is strongly decreased and the spatial quantization becomes important. The presence of opposite wavevectors in the quantized states averages out the RD SOC, and the observed pattern becomes that of the TE-TM SOC with four lobes exhibiting a double winding (Figure 4c). In real space (Figure 4d), the emission from the edges exhibits complicated circular polarization patterns due to the spatial superposition of quantum states.

**Conclusion**

In summary, we have observed room-temperature polariton condensation in organic single-crystal DPAVBi microbelts without any mirrors. Experimental and theoretical evidence shows that the photonic spin-orbit coupling in the anisotropic DPAVBi crystals leads to the optical spin Hall effect pattern switching in polariton condensation, which leads to the unidirectional propagation of left-right circularly polarized light on quantized states. Our work demonstrates the enormous potential of organic single crystals in the nonlinear and gain interaction properties towards novel topological polaritons systems.

**Acknowledgments:** The authors thank Dr. HW Yin from ideaoptics Inc. for the support on the angle-resolved spectroscopy measurements.

**Funding:** This work was supported by the National Natural Science Foundation of China (Grant No. 22150005, 22303023, 22090022, and 22275125), the National

Key R&D Program of China (2022YFA1204402, 2018YFA0704805 and 2018YFA0704802), the Natural Science Foundation of Beijing, China (KZ202110028043), the Hebei Education Department (Grant No. QN2024277); Advanced Talents Incubation Program of the Hebei University (521100222065), Beijing Advanced Innovation Center for Imaging Theory and Technology. This research was supported by the ANR Labex GaNext (ANR-11-LABX-0014), the ANR program "Investissements d'Avenir" through the IDEX-ISITE initiative 16-IDEX-0001 (CAP 20-25), the ANR project "NEWAVE" (ANR-21-CE24-0019) and the European Union's Horizon 2020 program, through a FET Open research and innovation action under the grant agreement No. 964770 (TopoLight).

**Author contributions:** J. Ren and Q. Liao developed the experiments and plans. J. Ren fabricated and characterized the samples. T. Long provided the equipment maintenance. D. Solnyshkov and G. Malpuech performed the theoretical calculations. All authors analyzed and discussed the results. J. Ren, C. Gu, H. Fu, D. Solnyshkov. G. Malpuech and Q. Liao prepared the manuscript with contributions from all the authors.

**Competing interests:** Authors declare that they have no competing interests.

**Data and materials availability:** All data are available in the main text or the supplementary materials.

**Correspondence:** Correspondence should be addressed to: dmitry.solnyshkov@uca.fr (D.S.), liaoqing@cnu.edu.cn (Q.L.)

**Supplementary Materials**

**Materials and Methods**

**Table S1** Coupled Harmonic Oscillator Model Fitting Results for $LP_1$ to $LP_{11}$.

**Figure S1.** a, b: Typical SEM image (a) and fluorescence image (b) of the pure DPAVBi crystal. c, d: SAED pattern measured by directing the electron beam to the top surface of a single microbelt shown in b. e: The molecular arrangement of the DPAVBi crystal.

**Figure S2.** The dispersion curve of uncoupled cavity modes of the pure DPAVBi microbelt and the corresponding fitting data (red dashed lines) using the planar-cavity-mode model. The effective refractive index obtained by the fitting is $n_{eff}$ = 2.15.

**Figure S3.** a: The experimentally measured dispersion relationship of the wavelength and $k_y$ wavevector ($k_x$ = 0) in unpolarized (a), σ+ (b) and σ- (c) circular polarization light at $P > P_{th}$ of DPAVBi microbelt.

# Supplementary Materials

# Quantum Spin Hall effect of topological polariton condensate from organic single-crystal microbelts

Jiahuan Ren,[1,2] Teng Long,[1] Chunling Gu,[3] Hongbing Fu,[1] Dmitry Solnyshkov,[4,5,*] Guillaume Malpuech,[4] Qing Liao[1,*]

[1]Beijing Key Laboratory for Optical Materials and Photonic Devices, Department of Chemistry, Capital Normal University, Beijing 100048, China

[2] College of Physics Science and Technology, Hebei University, Baoding 071002, China

[3]Institute of Process Engineering, Chinese Academy of Sciences, Beijing, 100190, China

[4]Institut Pascal, PHOTON-N2, Université Clermont Auvergne, CNRS, Clermont INP, F-63000 Clermont-Ferrand, France

[5]Institut Universitaire de France (IUF), 75231 Paris, France

**Materials and Methods**

**1. Preparation of the 2D microbelts of DPAVBi**

The organic molecule, 4,4′-bis[4-(di-p-tolylamino)styryl]biphenyl (DPAVBi, > 98%), was purchased from Aldrich and used directly without further purification.

The two-dimension (2D) microbelts of DPAVBi were prepared by using a solution reprecipitation method. Typically, 50 μL of a stock solution of DPAVBi in tetrahydrofuran (THF) was injected into 2 mL of hexane under stirring. The precipitates were centrifugally separated from the colloidal suspension and washed twice with hexane prior to vacuum drying.

**2. Structural and spectroscopic characterization**

As-prepared BDAVBi microbelts were characterized by field emission scanning electron microscopy (FE-SEM, HITACHI S-4800) by dropping on a silicon wafer. The X-ray diffraction (XRD, Japan Rigaku D/max-2500 rotation anode X-ray diffractometer, graphite monochromatized Cu Kα radiation ($\lambda$ =1.5418 Å)) operated in the 2θ range from 5 to 30°, by using the samples on a cleaned glass slide.

The fluorescence micrograph, diffused reflection absorption and emission spectra were measured on Olympus IX71, HITACHI U-3900H, and HITACHI F-4600 spectrophotometers, respectively.

**3. The angle-resolved spectroscopy characterization**

Note that in all spectral measurements in this work, the belt length was always projected to be parallel to the slit of the spectrometer.

The angle-revolved spectrum was measured using a Halogen lamp with wavelength range of 400-700 nm. The light source passes through a condenser with a numerical aperture of 0.9 and the effective illumination angle is ±50°. The linear polarizer was added to the incoming optical path to measure absorption spectrum of DPAVBi microbelts in different polarization directions.

The angle-revolved reflectivity was collected by using the same 100× microscope objective with a high numerical aperture (0.95 NA). The measurement angle can achieve ±71.8°. The reflectivity and photoluminescence spectroscopy were detected at room temperature in a home-made micro-area Fourier image which is presented to the spectrometer slit through four lenses. The reflectivity spectrum was collected by the spectrometer with a 300 lines/mm grating and a 400×1340 pixel liquid nitrogen cooled charge-coupled (CCD). For off-resonant optical pumping (400 nm, pulse width 150 fs) from a 1 kHz Ti: sapphire regenerative amplifier and 40-μm spot diameter with a near Gaussian beam profile, the angle-resolved photoluminescence was collected with the 1200 lines/mm grating to improve the resolution of the spectrometer.

In order to investigate the polarization properties, we placed a linear polarizer, a half-wave plate and a quarter-wave plate in front of spectrometer to obtain the polarization state of each pixel of the k-space images in the horizontal-vertical (0° and 90°), diagonal (±45°) and circular (σ+ and σ−) basis (S. Dufferwiel *et al.*, *Phys. Rev. Lett.* 2015, 115, 246401. & F. Manni *et al.*, *Nat. Commun.* 2013, 4, 2590.). Thereby, one can calculate the Stokes vector through:

$$S_1 = \frac{I_{0°} - I_{90°}}{I_{0°} + I_{90°}}$$

$$S_3 = \frac{I_{\sigma+} - I_{\sigma-}}{I_{\sigma+} + I_{\sigma-}}$$

## 4. Crystalline Characterization

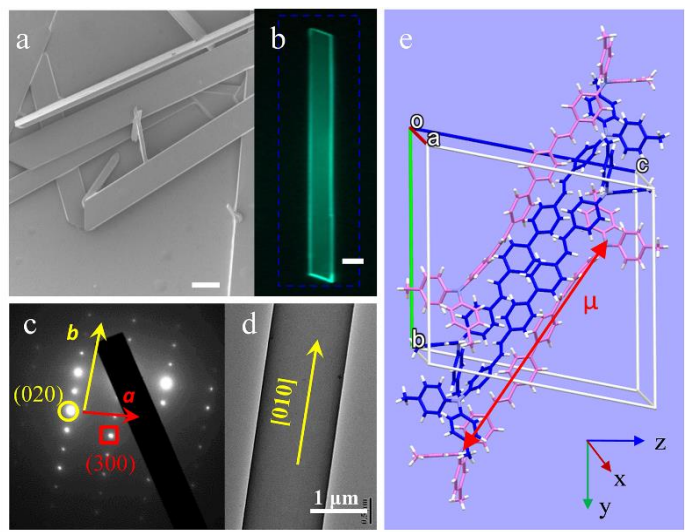

**Figure S1.** a, b: Typical SEM image (a) and fluorescence image (b) of the pure DPAVBi crystal. c, d: SAED pattern measured by directing the electron beam to the top surface of a single microbelt shown in b. e: The molecular arrangement of the DPAVBi crystal.

According to scanning electron microscopy (SEM) (Figure S1a), the obtained DPAVBi microstructures display the perfect belt-like shape with a typical width of 5-8 μm and a length of several hundreds of micrometers. Figure S1b presents the corresponding fluorescence microscopy of the selected DPAVBi microbelt, showing uniform and bright green-color emission. Selected area electron diffraction (SAED) pattern (Figure S1c) measured by directing the electron beam to the top surface of a single microbelt reveals that DPAVBi microbelts are single-crystals. According to the crystal database of DPAVBi triclinic crystalline form (CCDC: No. 1960793), these microbelts are grown along the crystal [010] direction (i.e. the crystal b-axis) and are bounded by (100) and ($\bar{1}$00) crystal planes on the lateral surfaces (Figure S1d). Figure 1e depicts the molecular arrangement with herringbone packing along the microbelt length direction. The transition dipole-moment (μ) of DPAVBi molecules (red double arrow line) along the molecular long-axis and tilts at an angle of 36° relative to the microbelt length direction. Thus, the projection of μ along the length direction (denoted as $k_y$ direction) much larger than the width direction (denoted as $k_x$ direction), which results in the giant anisotropic refractive index distribution in the DPAVBi microbelt.

**5. Polariton dispersion**

The polariton dispersion in Fig. 1b was calculated by a coupled harmonic oscillator Hamiltonian (CHO) model (S. Kena-Cohen *et al.*, *Phys. Rev. Lett.* 2008, 101, 116401.). The 2×2 matrix in equation (1) below describes the CHO Hamiltonian:

$$\begin{pmatrix} E_{CMn}(\theta) & \Omega/2 \\ \Omega/2 & E_X \end{pmatrix} \begin{pmatrix} \alpha \\ \beta \end{pmatrix} = E \begin{pmatrix} \alpha \\ \beta \end{pmatrix} \qquad (1)$$

Where $\theta$ represents the polar angle, $E_{CMn}(\theta)$ is the cavity photon energy of the $n^{th}$ cavity mode as a function of $\theta$, $E_X$ is the exciton 0–0 absorption energy of DPAVBi microbelts at 2.74 eV (452 nm) and $\Omega$ (eV) denotes the coupling. The magnitudes $|\alpha|^2$ and $|\beta|^2$ correspond to the photonic and the excitonic fraction, respectively.

The cavity photon dispersion is given by:

$$E_{CMn}(\theta) = \sqrt{\left(E_c^2 \times \left(1 - \frac{\sin^2\theta}{n_{eff}^2}\right)^{-1}\right)} - (n-1) \times l \qquad (2)$$

where $E_c$ represents the cavity modes energy at $\theta = 0°$, $E_{CM1}(\theta)$ represents the energy of the first cavity mode when n=1, $(n-1) \times l$ represents the energy difference from the first cavity mode. The effective refractive index ($n_{eff}$ = 1.6 and 2.15) is extracted from the fitting results. The theoretical fitting dispersion of the uncoupled cavity modes ($n_{eff}$ = 1.6, red solid line) and coupled cavity modes ($n_{eff}$ = 2.15, black dash line) is shown in Figure 1d, e. Diagonalization of this Hamiltonian yields the eigenvalues, $E_{\pm}(\theta)$, which represents the upper and lower polariton (UP and LP) in-plane dispersions (H. Deng *et al.*, *Rev. Mod. Phys.* 2010, 82, 1489-1537.),

$$E_{\pm}(\theta) = \frac{E_X + E_{CMn}(\theta)}{2} \pm \frac{1}{2}\sqrt{\left(E_X - E_{CMn}(\theta)\right)^2 + \hbar^2\Omega^2} \tag{3}$$

**Table S1.** Coupled Harmonic Oscillator Model Fitting Results for $LP_1$ to $LP_{11}$.

| Coupling mode | $LP_1$ | $LP_2$ | $LP_3$ | $LP_4$ | $LP_5$ | $LP_6$ | $LP_7$ | $LP_8$ | $LP_9$ | $LP_{10}$ | $LP_{11}$ |
|---|---|---|---|---|---|---|---|---|---|---|---|
| Detuning (meV) | 107 | 92 | 77 | 62 | 47 | 32 | 17 | 2 | -13 | -28 | -43 |
| Rabi splitting | | | | | | 390 meV | | | | | |
| $|\beta|^2$ | 0.74 | 0.71 | 0.68 | 0.65 | 0.62 | 0.58 | 0.54 | 0.51 | 0.47 | 0.43 | 0.39 |
| $|\alpha|^2$ | 0.26 | 0.29 | 0.32 | 0.35 | 0.38 | 0.42 | 0.46 | 0.49 | 0.53 | 0.57 | 0.61 |

Note: The magnitudes $|\alpha|^2$ and $|\beta|^2$ present the component of exciton and photon of the LPB, respectively.

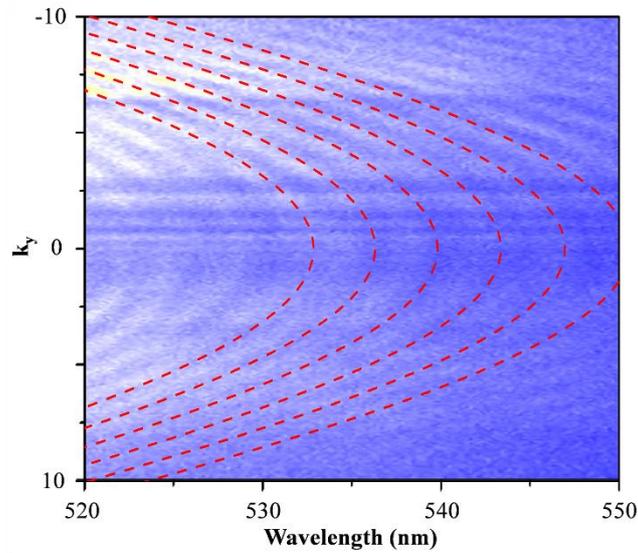

**Figure S2.** The dispersion curves of the pure DPAVBi microbelt in the spectral range (520-550 nm) and the corresponding fitting results (red dashed lines) using the planar-cavity-mode model. The effective refractive index obtained by the fitting is $n_{eff}$ = 2.15.

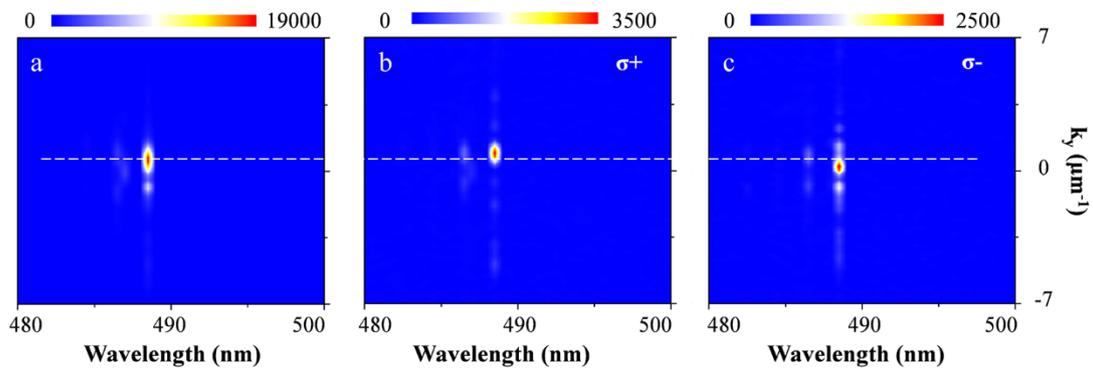

**Figure S3.** a: The experimentally measured dispersion relationship of the wavelength and $k_y$ wavevector ($k_x$ = 0) in unpolarized (a), σ+ (b) and σ- (c) circular polarization light at $P > P_{th}$ of DPAVBi microbelt.